# Indirect reciprocity can overcome free-rider problems on costly moral assessment

Tatsuya Sasaki[1]*, Isamu Okada[2], Yutaka Nakai[3]

[1]Faculty of Mathematics, University of Vienna, Oskar-Morgenstern-Platz 1, 1090 Vienna, Austria

[2]Department of Business Administration, Soka University, 1-236 Tangi-cho, Hachioji-city, Tokyo, 192-8577, Japan

[3]Faculty of Systems Engineering and Science, Shibaura Institute of Technology, Fukasaku 2307, Minuma-ku, Saitama-city, Saitama, 337-8570, Japan

*Correspondence to: tatsuya.sasaki@univie.ac.at

Revised on May 25, 2016

**Abstract:** Indirect reciprocity is one of the major mechanisms of the evolution of cooperation. Because constant monitoring and accurate evaluation in moral assessments tend to be costly, indirect reciprocity can be exploited by cost evaders. A recent study crucially showed that a cooperative state achieved by indirect reciprocators is easily destabilized by cost evaders in the case with no supportive mechanism. Here, we present a simple and widely applicable solution that considers pre-assessment of cost evaders. In the pre-assessment, those who fail to pay for costly assessment systems are assigned a nasty image that leads to being rejected by discriminators. We demonstrate that considering the pre-assessment can crucially stabilize reciprocal cooperation for a broad range of indirect reciprocity models. In particular for the most leading social norms we analyse the conditions under which a prosocial state becomes locally stable.

# Introduction

Natural selection disfavours indirect reciprocity unless the costs of observation and assessment are negligible [1]. According to proper social norms that distinguish good from evil, such as image-scoring norm [2], indirect reciprocity can promote cooperation even in large populations [3]. However, making moral assessments takes time and effort. Discriminators who incur no assessment cost thus appear as free riders that erode a cooperative state achieved by discriminators who incur the costs. Despite the advance of indirect reciprocity, the crucial question remains unsolved [4]: How can cooperation through indirect reciprocity be maintained when considering the costs associated with the assessment system?

To address this crucial question, we focus on the fact that this puzzling situation is closely related to the second-order free-rider problem in costly punishment [1]. The evolution of costly punishment, in striking contrast to indirect reciprocity, has been given much more attention over the last decades. In tackling the second-order free-rider problem, previous study significantly examined pool punishment [5-10]. The key aspect of pool punishment is its proactive mechanism to detect second-order free riders through unconditional pre-payment. The mechanism paves the way for effectively punishing second-order free riders.

In this paper we apply the essence of the pool-punishment mechanism to fix the issue of the costly moral assessment. In the next section, we introduce a basic model of indirect reciprocity and the known negative outcome from considering the assessment costs. In the Results section we show how adopting a proactive assessment mechanism can improve the outcome.

**Materials and methods**

We build upon the standard framework for the evolution of indirect reciprocity by reputation [11,12]. Using the framework, a strategy for discriminators is given by an assessment rule combined with an action rule. We base indirect reciprocity on the giving game, which is a two-player game in which one player acts as a donor and the other a recipient. The donor can choose to help the recipient by giving benefits $b>0$ at personal cost $c>0$ or not to help. We consider the following implementation error: a player who has intended to help involuntarily fails to do so with a probability $e$ [1,13].

We start with a basic model in which each individual is endowed with a binary image score of 'good' or 'bad'. It is assumed that the discriminator's action rule is to help a good recipient or not to help a bad recipient. After observing every giving game, a unique assessment system assigns the donor's image by following a specific assessment rule. We assume that all discriminators share the same list of individual image scores provided by the assessment system. We later consider in particular the second-order assessment rule, which is a function of the donor's last action and the recipient's last image (Table 1).

To study the evolution of discriminators, we respect a continuous-entry model: an individual's birth and death sometimes happen, and this changes the strategy distribution in the population [14]. We assume that in one's lifetime an individual infinitely plays the one-round giving game with different opponents. We consider infinitely large populations and analyse the replicator dynamics [15] for the following four strategies: 1) Paying discriminator [Z] is willing to help a good recipient and refuses to help a bad recipient in the giving game. Also s/he is willing to pay for the assessment cost $k>0$. 2) Evading discriminator [W] similarly acts as a paying discriminator in the giving game, except that s/he is not willing to pay for the assessment cost. 3) Cooperator [X] unconditionally intends to help a potential recipient, and 4) Defector [Y] unconditionally intends not to help a potential recipient. Both cooperator and defector are not willing to pay for the assessment cost. We denote by $x$, $y$, $z$ and $w$ the frequencies of cooperators, defectors and paying and evading discriminators, respectively. The replicator dynamics for these strategies are described as $dn/dt = n(P_S - P)$, where $n$ is the frequency of strategy $S$ (= X, Y, Z, W), $P_S$ is the expected payoff given by the limit in the mean of the payoff per round for strategy $S$ and $P$ is the average payoff over the population, given by $xP_X + yP_Y + zP_Z + wP_W$.

To formalize the expected payoffs, we denote by $g_S$ the probability that a recipient with strategy $S$ is helped by a given discriminator. In the basic model this is identical to the fraction of good players within all $S$ strategists. Let $g$ be the population average of $g_S$, thus $g = xg_X + yg_Y + zg_Z + wg_W$. The population size is very large, so we may assume that the population configuration for $g_S$ does not change between the consecutive one-round giving games [16]. Thus, the expected payoffs are described as

$$\begin{aligned} P_X &= (1-e)b[x + g_X(z+w)] - (1-e)c, \\ P_Y &= (1-e)b[x + g_Y(z+w)], \\ P_Z &= (1-e)b[x + g_Z(z+w)] - (1-e)cg - k, \\ P_W &= (1-e)b[x + g_W(z+w)] - (1-e)cg. \end{aligned} \quad (1)$$

We note that in the basic model either paying or evading discriminators intend to help a potential recipient who has a good image, thus leading to $g_Z = g_W$. This results in paying discriminators being worse off than evading discriminators. Substituting this into Eq. (1) yields

$$P_Z - P_W = -k < 0. \quad (2)$$

For any degree of the assessment cost $k$, evading discriminators dominate paying discriminators in the interior state space, and thus the population in the end attains a state that excludes paying discriminators. In the absence of cost payers, the assessment system cannot be established. Consequently, cooperation in that case would vanish without discrimination.

**Results**

The basic model reveals that considering cost evaders destabilizes indirect reciprocity irrespective of the assessment rule, as shown in previous work [1]. To stabilize indirect reciprocity, we examine an institutional variant of the basic model. As a first step, we extend the basic model to a two-stage game in which one round consists of the stage of payment for the observation costs followed by the stage of the giving game, which is the same as in the basic model. The first stage offers an opportunity to transfer some fees to a central account as in automatic utility payments.

The essential idea is to specifically assess the second-order free rider. We consider a different binary moral code 'nice' or 'nasty'. The (unique) assessment system assigns a nice image to an individual if s/he pays the costs in the first stage, otherwise that individual is assigned a nasty image. In evaluating the donor's action of the giving game, as the first step we simply apply the existing assessment framework to the second stage, as in the basic model.

We keep the four strategies, cooperators, defectors, paying discriminators and evading discriminators, as before and assume that in the first stage, paying discriminators are willing to pay but the remaining cooperators, defectors and evading discriminators are not. We also modify the discriminator's action rule for the giving game as follows: either paying or evading discriminators give help if a potential recipient has a good *and* nice image, or otherwise (if bad *or* nasty), refuse help.

The extra assessment by the utilities payment system seriously lowers the image score for second-order free riders. For analytical simplicity, we assume that the utilities payment system is so perfect that no assessment error occurs for the first stage. All of evading strategies: cooperators, defectors and evading discriminators (X, Y and W), therefore, are necessarily assessed as nasty. This yields $g_X = g_Y = g_W = 0$. (Note that in the variant, $g_S$ equals the probability of good and nice players.) Thus, Eq. (1) becomes

$$
\begin{aligned}
P_X &= (1-e)bx - (1-e)c, \\
P_Y &= (1-e)bx, \\
P_Z &= (1-e)b[x + g_Z(z+w)] - (1-e)cg_Z z - k, \\
P_W &= (1-e)bx - (1-e)cg_Z z.
\end{aligned}
\quad (3)
$$

It is clear that $P_Y \geq P_W \geq P_X$. To understand when the homogeneous state of paying discriminators $z = 1$ becomes locally stable, it is enough to check if $P_Z - P_Y > 0$ in the vicinity of $z = 1$ on the face $x = w = 0$. This yields

$$P_Z - P_Y = (1-e)(b-c)g_Z z - k. \quad (4)$$

With suitable assessment rules, it is possible to have that $g_Z > 0$ in the vicinity of $z = 1$. In this case, the node $z = 1$ turns into a locally stable equilibrium when the net benefit $b - c$ is sufficiently large compared to the assessment cost $k$.

Finally, we demonstrate how the variant improves the results for some of the most leading assessment rules. We examine simple standing [13,16] and stern judging [17], the only two second-order assessment rules in the leading eight norms [10,11]. According to the discriminator's action rule in the variant, we extend simple standing and stern judging as in table 1. These rules assign a good image to those who help a good and nice recipient with no implementation error (probability $(1-e)g$) and also a good image to those who refuse to help a bad or nasty recipient (probability $1-g$). By assumption of the image dynamics, the sum of these probabilities should equal $g_Z$. Considering also $g = g_Z z$ then leads to the recursive equation for $g_Z$, $g_Z = (1-e)g_Z z + (1-g_Z z)$. This yields $g_Z = 1/(1+ez)$. Hence, the necessary and sufficient condition for the homogeneous state of paying discriminators ($z = 1$) to be locally stable either under the simple-standing or stern-judging rule is

$$\frac{1-e}{1+e}(b-c) - k > 0. \tag{5}$$

Figure 1a shows the basin of attraction for $z = 1$, and figure 1b depicts the flow on the boundary faces of the state space under simple standing. If we assume assessment errors in the first stage, the image dynamics become more complicated but the main results remain qualitatively unchanged—paying discriminators can stabilise with the pre-assessment of cost evaders (electronic supplementary material, S1).

**Discussion**

Since the definitive 2013 work by Suzuki and Kimura [1], the evolution of indirect reciprocity relying on costly assessment systems has been explicitly recognized as one of the inevitable issues that challenge the advance of indirect reciprocity [4]. To address the issue, we considered a simple pre-assessment mechanism that is set prior to the primary game in order to detect and label cost evaders. We then demonstrated that the mechanism considered leads to stabilizing costly indirect reciprocity under the most leading social norms, simple standing and stern judging.

Our results are potentially applicable to a broad range of existing indirect reciprocity models, such as tolerant scoring [18], group scoring [19], reputation-based punishment [20], mixed public and private interactions [21], optional interactions [22] and finite populations [23]. On the one hand, managing more complicated assessment systems, such as in [18-21], would be more costly, and thus it is worth considering pre-assessment mechanisms for reducing the temptation to evade cost sharing. On the other hand, as in the case of pool punishment [8], jointly considering optional interactions [22] and finite populations [23] might facilitate establishing pre-assessment mechanisms.

Another promising avenue for future studies would be to explore costly indirect reciprocity on more realistic structured populations. Recent studies using structured populations suggest the importance of cooperator assortment based on reputation [24, 25]. However, little is known about how information cost affects reputation-based reciprocity on a network. In the case of the second-order free-rider problem in costly punishment, considering the locality of interactions among players can solve the problem by separating costly punishers from second-order free riders [26]. Similarly, the extension to structured populations may lead to significantly different outcomes for paying and evading discriminators.

We left out an advanced issue of analysing nonlinking discriminators [27] who act as paying discriminators yet are willing to help cost evaders with a good image. Nonlinking discriminators can invade paying discriminators by neutral drift. The preliminary results indicate that considering implementation or assessment errors for the first stage can lead paying discriminators to become better off than nonlinking ones, as in fixing neutral drift between conditional and unconditional cooperators [13]. Further investigation is planned in future work.

We note that prepayment for assessment systems can be viewed as a kind of contribution to collective action. Thus, our results corroborate those of previous studies on two-stage games in which reciprocal behaviours in the second stage are linked to a collective action in the first stage. For instance, Panchanathan and Boyd showed that collective action in the first stage can be maintained by considering a shunning strategy that in the second stage withholds help for those who failed to contribute in the first stage [28]. Together, the present results further imply that such a proactive social mechanism that can discriminate those who deserve to enter social exchange and reciprocal norms within social exchange may evolve jointly.

**Acknowledgements.** T.S. acknowledges the Austrian Science Fund (FWF): P27018-G11. I.O. acknowledges Grant-in-Aid for Scientific Research (B) 16H03120. Y.N. acknowledges Grant-in-Aid for Scientific Research (B) 16H03698.

**Authors' contributions.** T.S., I.O. and Y.N. designed and analysed the model, and T.S. wrote the paper.

**Table caption**

**Table 1.** How second-order rules make moral assessments in giving games with pre-assessment. 'G' and 'B' describe a good and bad image, respectively. In the donor's action, 'C' and 'D' describe giving help and refusing help, respectively.

**Figure caption**

**Figure 1.** Pre-assessment of cost evaders stabilizes costly indirect reciprocity. (*a*) The tetrahedron describes a simplex of the state space $\{(x,y,z,w): x+y+z+w=1; x,y,z,w \geq 0\}$. Each corner corresponds to the homogeneous state of each specific strategy. The basin of attraction for paying discriminators covers approx. 61.5% of the whole space. (*b*) The flow diagrams depict the direction of evolution on the boundary faces of the state-space simplex. The state space has no interior equilibrium, and all interior orbits converge to the boundary. Any mixed state of defectors and evading discriminators forms an equilibrium point on the edge $y+w=1$. Parameters: $c=1$, $b=1.5$, $e=0.01$, $k=0.3$, and simple-standing rule.

**Table**

| Conditions | Recipient's image | G and nice | G and nice | B or nasty | B or nasty |
|---|---|---|---|---|---|
| | Donor's action | C | D | C | D |
| Assessment rule: What does donor's image look like? | Simple standing | G | B | G | G |
| | Stern judging | G | B | B | G |

# Figure

(a)

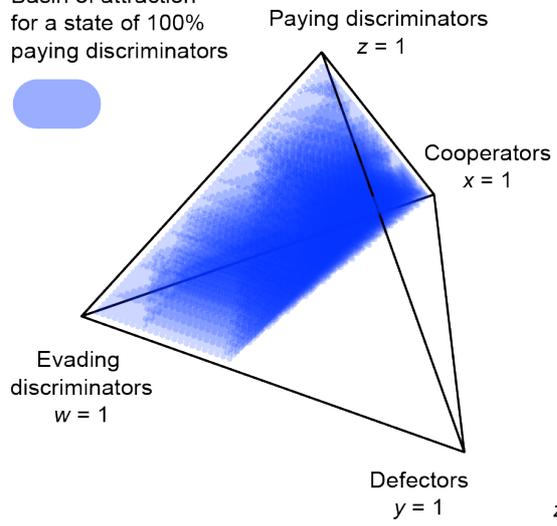

Basin of attraction for a state of 100% paying discriminators

Paying discriminators $z = 1$

Cooperators $x = 1$

Evading discriminators $w = 1$

Defectors $y = 1$

(b)

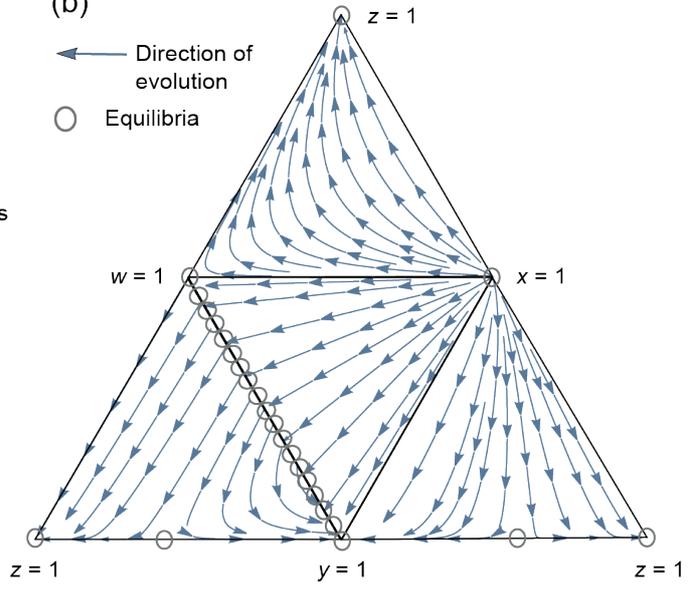

← Direction of evolution

○ Equilibria

$z = 1$

$w = 1$

$x = 1$

$z = 1$

$y = 1$

$z = 1$

# Electronic supplementary material for:
# Indirect reciprocity can overcome free-rider problems
# on costly moral assessment

Tatsuya Sasaki, Isamu Okada, Yutaka Nakai

## S1. Simple standing and stern judging with first-stage assessment errors

We examine assessment errors for the first stage of the variant model considered in the main text. We explore the conditions for the homogeneous state of paying discriminators [Z] with $z = 1$ to become stable under simple standing and stern judging. We check when paying discriminators become better off than the other three strategies: cooperators, defectors and evading discriminators.

First, we analyse the frequency of good and nice players among $S$-strategists (S = X, Y, Z or W), $g_S$. As in the main text, we assume that the degree of $g_S$ is unchanged between the consecutive one-round (two-stage) games. We note that by definition the only difference between the rules is with respect to how a potential donor is to be assessed when a potential recipient is bad or nasty and the donor's action is not to help, in which case simple standing assigns a good image and stern judging a bad image.

We denote by $e_1$ the probability of a first-stage assessment error in which the assessment system involuntarily assesses a paying player (who should have been nice) as nasty or an evading player (who should have been nasty) as nice.

For simple standing, $g_S$ is given by

$$\begin{aligned}
g_X &= e_1[(1-e_2)g + (1-g)], \\
g_Y &= e_1[0 \cdot g + (1-g)], \\
g_Z &= (1-e_1)[(1-e_2)g + (1-g)], \\
g_W &= e_1[(1-e_2)g + (1-g)],
\end{aligned} \qquad (S1)$$

in which $g$ denotes the frequency of players who have both a good and nice image over the whole population, thus $g = xg_X + yg_Y + zg_Z + wg_W$, and $e_2$ describes the probability of an implementation error in the second stage (see the main text for details). In equation (S1), the evading players (with X, Y or W) and the paying players (with Z) are assessed as nice with probability $e_1$ and $1-e_1$, respectively, in the first stage. In addition, the bracket terms of the right side describe the probability that a donor with strategy $S$ is assessed as good in the second-stage giving game. When a recipient has a good and nice image (with probability $g$), X, Z and W strategists are willing to help and are assessed as good with probability $(1-e_2)g$ and Y strategists refuse to help, thus receiving a bad image. Simple standing is a tolerant norm, which is to assign a good image to a donor, irrespective of his/her actions to a recipient who has a bad or nasty image. This leads to the same second term in the bracket as $1-g$ over all $g_S$.

Then, for stern judging, equation (S1) becomes

$$g_X = e_1[(1-e_2)g + e_2(1-g)],$$
$$g_Y = e_1[0 \cdot g + (1-g)],$$
$$g_Z = (1-e_1)[(1-e_2)g + (1-g)],$$
$$g_W = e_1[(1-e_2)g + (1-g)].$$
(S2)

Stern judging assigns a good image to those who refuse to help a bad or nasty recipient and a bad image to those who help a bad or nasty recipient. This leads to the second term in the bracket for $g_X$, $e_2(1-g)$, which is the only difference from simple standing in equation (S1).

We analyse the expected payoff $P_S$ at the point $z=1$. Equation (1) in the main text is specified as:

$$P_X = (1-e_2)bg_X - (1-e_2)c,$$
$$P_Y = (1-e_2)bg_Y,$$
$$P_Z = (1-e_2)bg_Z - (1-e_2)cg - k,$$
$$P_W = (1-e_2)bg_W - (1-e_2)cg.$$
(S3)

Considering equations (S1) to (S3), it is obvious that $P_W$ is greater than or equal to $P_X$. Thus, if $P_Z - P_W\big|_{z=1} > 0$ holds, this yields $P_Z - P_X\big|_{z=1} > 0$.

By solving equations (S1) and (S2) for $z = 1$, we obtain, in either case of simple standing or stern judging,

$$g\big|_{z=1} = g_Z\big|_{z=1} = \frac{1-e_1}{1+e_2(1-e_1)}. \tag{S4}$$

Substituting equation (S4) into equations (S1) to (S3) yields

$$P_Z - P_Y\big|_{z=1} = \frac{b(1-e_2)[1-(1+e_2)e_1-(1-e_2)e_1^2]-c(1-e_1)(1-e_2)}{1+e_2(1-e_1)} - k, \tag{S5}$$

and

$$P_Z - P_W\big|_{z=1} = \frac{b(1-e_2)(1-2e_1)}{1+e_2(1-e_1)} - k. \tag{S6}$$

Equations (S5) and (S6) (which give the stability threshold conditions) for $z=1$ are common throughout simple standing and stern judging.

As the degree of the first-stage assessment error $e_1$ goes to 0, equations (S5) and (S6) converge to $\frac{1-e_2}{1+e_2}(b-c)-k$ and $\frac{1-e_2}{1+e_2}b-k$, respectively. Therefore, equation (5) in the main text,

$$\frac{1-e_2}{1+e_2}(b-c)-k > 0,$$

is sufficient for $z=1$ to also be stable for a sufficiently small degree of $e_1$ in either case of simple standing or stern judging.